\documentclass[aps,pre,twocolumn,groupedaddress,showpacs,superscriptaddress,amssymb,amsmath]{revtex4-1}
\usepackage{graphicx}
\usepackage{cancel}
\usepackage{tabularx}
\usepackage{color}
\usepackage{amsmath}
\usepackage{comment}
\newcommand{\be}{\begin{equation}}
\newcommand{\ee}{\end{equation}}
\newcommand{\bea}{\begin{eqnarray}}
\newcommand{\eea}{\end{eqnarray}}
\usepackage{dcolumn}
\usepackage{hyperref}
\usepackage{bm}
\usepackage{epsf}
\usepackage{subfig}
\usepackage{epstopdf}%
\usepackage{amsfonts}

\newcommand{\old}{\color{black}}
\let\la\langle \let\ra\rangle

\begin{document}

\title{Random walk with horizontal and cyclic currents}
\author{Joanna Li}
\altaffiliation{These authors contributed equally to this work.}
\affiliation{Department of Physics, University of Toronto, 60 Saint George St., Toronto, Ontario M5S 1A7, Canada}
\affiliation{Division of Engineering Science, University of Toronto, 42 Saint George St., Toronto, Ontario M5S 2E4, Canada}

\author{Matthew Gerry}
\altaffiliation{These authors contributed equally to this work.}
\affiliation{Department of Physics, University of Toronto, 60 Saint George St., Toronto, Ontario M5S 1A7, Canada}

\author{Israel Klich}
\affiliation{Department of Physics, University of Virginia, Charlottesville, Virginia 22903, USA}

\author{Dvira Segal}
\email{dvira.segal@utoronto.ca}
\affiliation{Department of Chemistry and Centre for Quantum Information and Quantum Control,
University of Toronto, 80 Saint George St., Toronto, Ontario, M5S 3H6, Canada}
\affiliation{Department of Physics, University of Toronto, 60 Saint George St., Toronto, Ontario M5S 1A7, Canada}


\date{\today}

\begin{abstract}
We construct a minimal two-chain random walk model and study the information that fluctuations of the flux and higher cumulants can reveal about the model: its structure, parameters, and whether it operates under nonequilibrium conditions.
The two coupled chains allow for both horizontal and cyclic transport. We capture these processes by deriving the cumulant generating function of the system, which characterizes both horizontal and cyclic transport in the long time limit.
First, we show that either the horizontal or the cyclic currents, along with their higher-order cumulants, can be used to unravel the intrinsic structure and parameters of the model. 
Second, we investigate the ``zero current" situation, in which the {\it horizontal} current vanishes. We find that fluctuations of the horizontal current reveal the nonequilibrium condition at intermediate bias, while the cyclic current remains nonzero throughout. We also show that in nonequilibrium scenarios close to the zero {\it horizontal} current limit, the entropy production rate is more tightly lower-bounded by the relative noise of the {\it cyclic} current, and vice versa. 
Finally, simulations of transport before the steady state sets in allow for the extraction of the interchain hopping rate.
Our study, illustrating the information concealed in fluctuations, could see applications in chemical networks, cellular processes, and charge and energy transport materials. 
\end{abstract}
\maketitle

\section{Introduction}

A common approach towards understanding stochastic processes, such as chemical reactions, biomolecular processes, or transport of charge or energy carriers, consists of representing them as Markov jump processes on coarse-grained networks of states \cite{vankampen}. This provides a framework for probing the first passage statistics \cite{zilman_first_passage, Kolom13} of such processes, as well as thermodynamic considerations \cite{udoCRN, rao2016, brown2017, rao2018}. In simple cases, networks are unicyclic, or one-dimensional, such that the states can be ordered unambiguously, and the system may only proceed through states in this order (albeit forward or reverse). Unicyclic networks are characterized by a single current, or flux, at steady state, whose statistics may be examined to determine the underlying properties of the network \cite{udo15, modular}, and whether the system is at equilibrium. Reaction networks may, in general, however, consist of multiple cycles, allowing for multiple fluxes to characterize the behavior of the system. 

When this is the case, an observer may have access only to partial information about the jump process playing out at the level of the reaction network. For example, groups of the underlying states may be combined into ``metastates", such that transitions within a metastate cannot be detected, obscuring information about currents. Numerous studies have assessed how partial information of this sort can be utilized to determine the rate at which entropy is produced in a system, characterizing deviations from thermal equilibrium \cite{skinner1, skinner2, martinez2019, harunari2022, nitzan2023, ghosal2023}. A related question pertains to what information about the structure of the underlying network itself can be determined through observations of only the observable currents. For example, one may consider whether the statistics exhibited by an observed current (its mean value as well as higher-order cumulants) can indicate that the underlying network is multicyclic, and that there may, therefore, be additional, hidden currents present \cite{vandermeer2022}.

One way of estimating entropy production based on partial information about fluxes in a system described by a reaction network is via the lower bound the thermodynamic uncertainty relation (TUR) places on this quantity relative to the precision of each current \cite{udoTUR, ging16, ging17, li2019, horowitz2020, bijayTUR_hierarchy, kamijima2023}. Since each independent current leads to an independent TUR bound, it is generally necessary to consider all currents to determine the tightest available bound on the entropy production. Having only partial information can therefore lead to issues in estimating the entropy production in this manner; the flux to which the observer has access may not be the one associated with the tightest TUR bound, as is certainly the case when the observable flux vanishes.

As a case of interest for systems with multiple fluxes, one may consider the situation in which an observed current stalls--that is, its mean value is zero. In general, this can occur away from equilibrium, and other, hidden fluxes, may take on nonzero mean values reflecting the nonequilibrium nature of the setup. Furthermore, higher-order cumulants of the stalling current generally deviate from equilibrium behavior, provoking questions concerning whether driving a system away from equilibrium can lead to advantageous effects (e.g., suppression of noise), even in the absence of a net flux. Past studies have worked to characterize and bound noise in cases like this for quantum charge and energy transport setups \cite{DeltaT18, JaninePRL, JaninePRB}.  We also note that in the quantum context, fluctuations have been used to characterize fractional charge \cite{de1998direct}, as well as to characterize the many-body entanglement of quantum states \cite{2006PhRvA..74c2306K,Klich:2008un,2010PhRvB..82a2405S}.  \old


In this study, we construct a minimal model for a system exhibiting multiple currents, amounting to a pair of one-dimensional, continuous-time random walks, between which the system can jump. The resulting system is characterized by two independent fluxes. The model allows some ambiguity as to exactly how the pair of fluxes are defined; we consider a {\it symmetric}
current tracking the 
horizontal
position of the walker along the pair of chains (in analogy to a simple random walk along a one-dimensional chain), and an {\it asymmetric} 
current reflecting the differences in how the two chains facilitate a walker's horizontal motion. The asymmetric current captures  
loops that can form in a walker's trajectory as it switches between the chains and begins to proceed in the reverse direction before switching back  (i.e., tracking the net number of steps the walker has taken in the clockwise direction).  In many cases, the asymmetric current can be considered as a {\it cyclic} current.
The coupled two-chain model serves as a useful tool for investigating what partial flux information can expose about the underlying structure of the network, as well as the deviation from equilibrium, and for examining the behavior of zero-current noise. The model is also useful in demonstrating how the distinct fluxes are unequal in their ability to set a lower bound for the entropy production, when there is available information about both.

In Sec. \ref{sec:model}, we introduce the random walk model in greater detail and outline the method of full counting statistics we use to calculate the fluxes and their fluctuations. Sec. \ref{sec:cumulants} focuses on the manners in which information about the underlying structure of the network are reflected in the cumulants of only a single current. In Sec. \ref{sec:zc_and_entropy}, we address the phenomenon of zero-current noise in the context of our model, elucidating the behavior of the higher-order cumulants of a stalling current in the situation where that current vanishes. We also compare the effectiveness with which the precision of each current can provide a bound on the entropy production rate, reflecting nonequilibrium behavior. In Sec. \ref{sec:dynamics}, we consider how finite-time observations of the dynamics can similarly be used to ascertain information about network structure. Finally, we summarize and look to future directions of study in \ref{sec:summary}.

\begin{figure}%
    \centering
    \subfloat[\centering Two chains biased in the same direction (i.e. to the right).] 
    {{\includegraphics[width=0.4\textwidth]{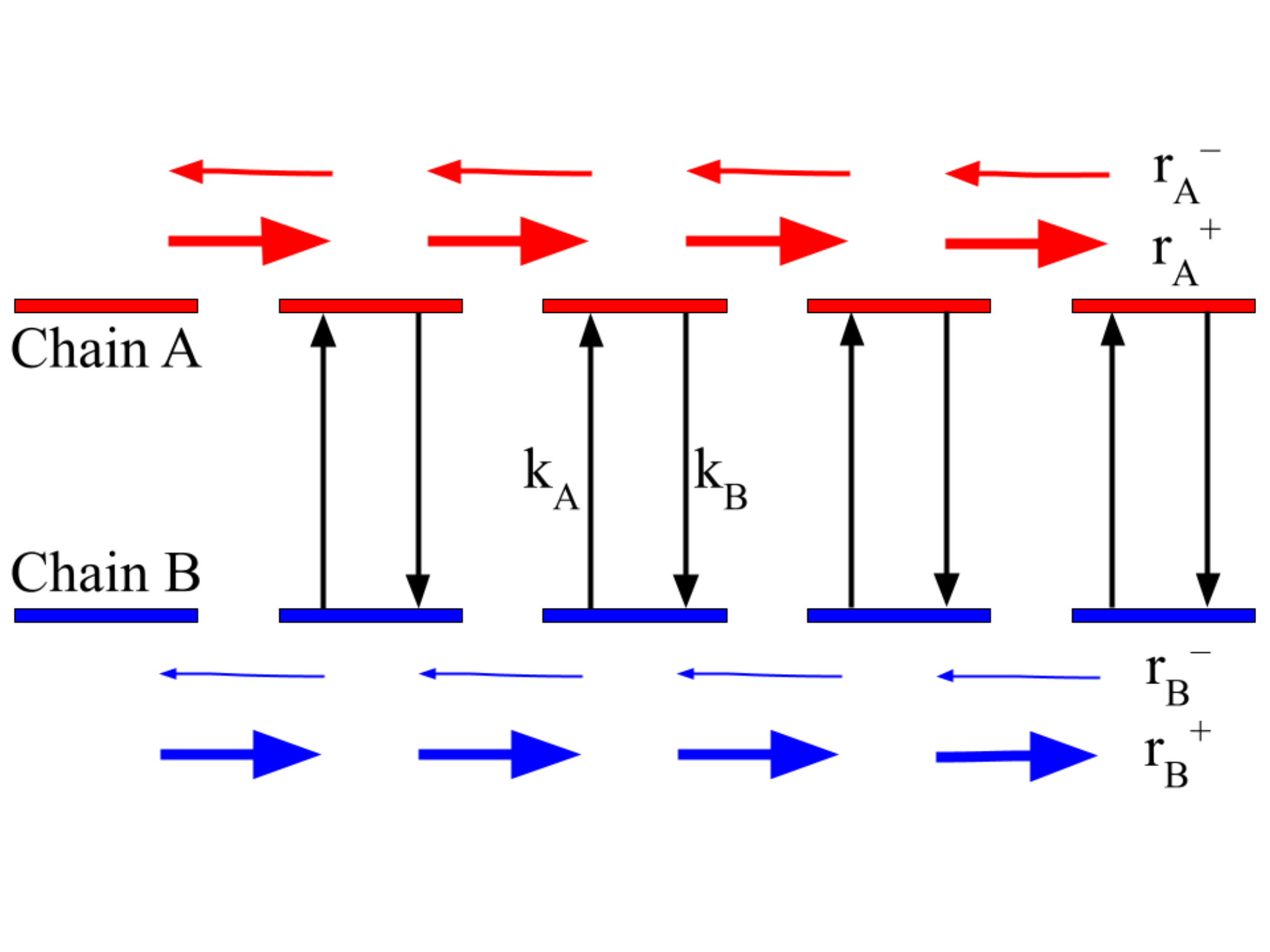}
    \label{fig:diffbias}}}
    \qquad
    \subfloat[\centering Two chains biased in opposite directions.]
    {{\includegraphics[width=0.4\textwidth]{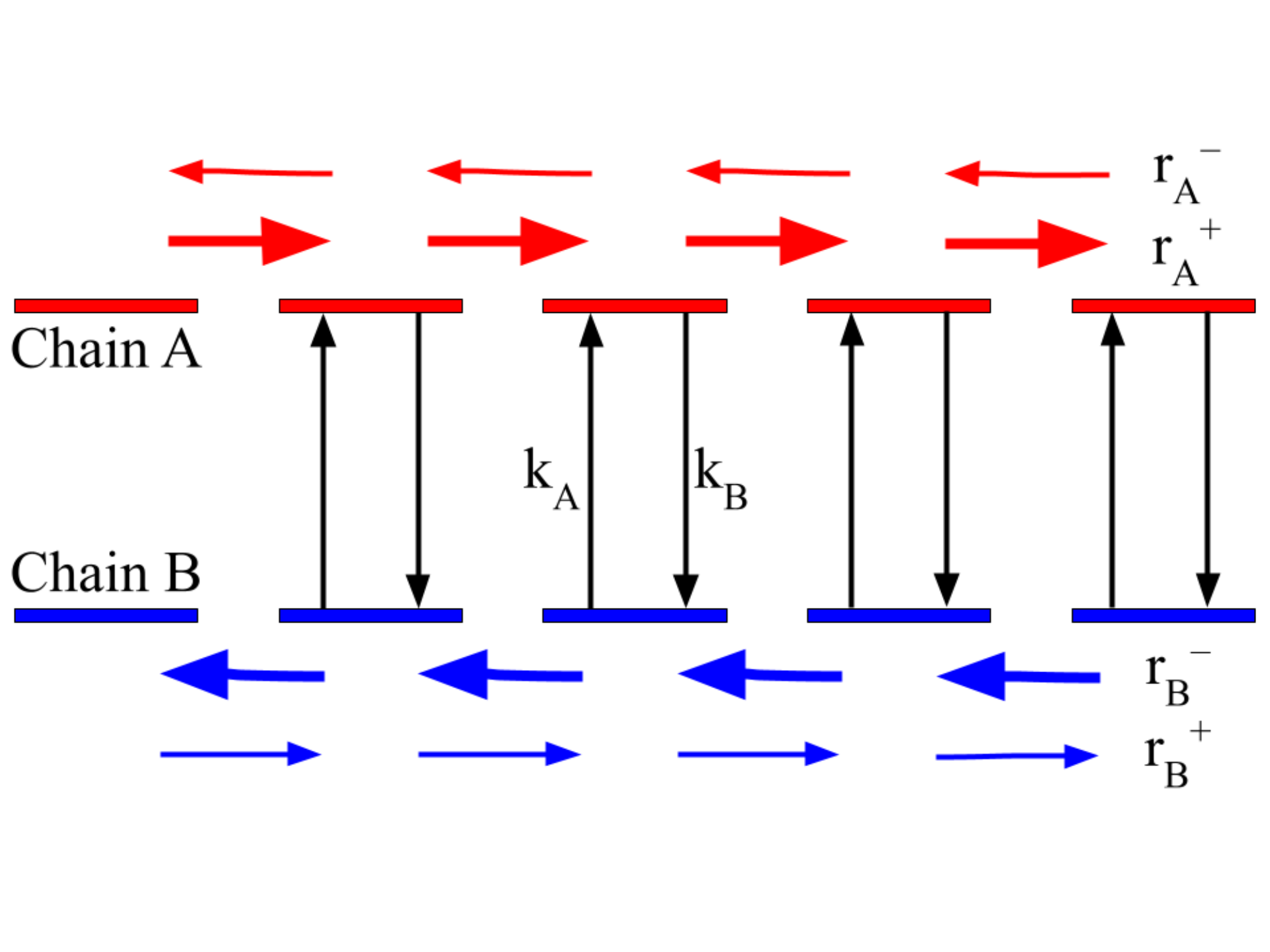}
    \label{fig:oppbias}}}
    \caption{Schematics of a two-chain random walk with six different transition rate constants, $r_A^{\pm}$, $r_B^{\pm}$, $k_A$ and $k_B$. Note that $k_A$ and $k_B$ are the same for every pair of states. The thicknesses of the arrows indicate the relative magnitude of transition rate constants. 
    }%
    \label{fig:model}%
\end{figure}

\section{Model and Method}
\label{sec:model}

\subsection{Model}
In a random walk with two currents, both currents provide useful information regarding the model's underlying structure. When one current is 
inaccessible, we can use the second current to deduce the properties of the system. To study this effect, we construct a minimal model of two distinct currents by coupling two one-dimensional chains denoted $A$ and $B$ (Fig. \ref{fig:model}). The two chains have forward and backward transition rate constants of $r_\nu^+$ and $r_\nu^-$, respectively, where $\nu = A, B$. Depending on their transition rate constants, biases along the two chains, that is, the difference between forward and reverse rates, may differ in magnitude (Fig. \ref{fig:diffbias}) and direction (Fig. \ref{fig:oppbias}). A walker can switch between the two chains
without changing its horizontal position,
with a rate constant of $k_A$ to enter chain $A$ from chain $B$, and $k_B$ to enter chain $B$ from chain $A$.

This general model has been used as a coarse-grained representation of a bacterial flagellar motor that incorporates directional switching \cite{korobkova2006,wang2014,skinner1}.
It may similarly be utilized in studying Brownian particles switching between diffusivities \cite{bressloff2020}. 
Additionally, it is related to the flashing ratchet model, where transitioning between the two chains is analogous to the switching off and on of an asymmetric potential, allowing, respectively, free and biased diffusion of the Brownian motor \cite{astumian1994,harms1997,astumian1997,ait-haddou2003,hwang2019}. This model is also broadly related to coarse-grained models of molecular machines, if the horizontal position along the chain is taken to represent a number of steps taken or the number of times a functional cycle has been completed, while the switching between the two chains may correspond to switching between (possibly hidden) internal states \cite{lau2007,teza2020,nitzan2023}. For instance, analogous models have been employed to analyze enzyme kinetics characterized by both a reaction and conformation coordinate \cite{cao1, cao2, cao3}.


We characterize the state of the system using a description consisting of three elements: which of the two chains ($A$ or $B$) the system is on, the  integrated symmetric current tracking,  the net number of steps, $n_h$, taken along the horizontal direction on either chain (i.e. horizontal distance from the origin), and the \textit{difference}, $n_c$, between the net number of steps taken on chain $A$ and those taken on chain $B$. This last quantity can be understood to be  the integrated asymmetric 
current. We refer to this also as the cyclic current, as it tracks the net number of steps the walker has taken in the clockwise direction, capturing cycles when they may arise\old. Steps towards the right on chain $A$ add to this total, while steps taken to the right on chain $B$ detract from it.

The resolved equations of motion for the probability distribution, over all possible states are,
\begin{align}\label{eq:ME}
    \dot{P}_A&(n_h,n_c;t) = \nonumber\\
     &-P_A(n_h,n_c;t)(r_A^+ + r_A^- + k_B) + P_B(n_h,n_c;t)k_A\nonumber\\
    & + P_A(n_h-1, n_c-1;t)r_A^+ + P_A(n_h+1,n_c+1;t)r_A^- \nonumber\\
    \dot{P}_B&(n_h,n_c;t) =\nonumber\\
     &-P_B(n_h,n_c;t)(r_B^+ + r_B^- + k_A)+ P_A(n_h,n_c;t)k_B\nonumber\\
    &+ P_B(n_h-1, n_c+1;t)r_B^+ + P_A(n_h+1,n_c-1;t)r_B^-.
\end{align}
%

\subsection{Full Counting Statistics}

We wish to investigate how the properties of the probability distributions reflect the features of the underlying system. Towards this goal, we derive the moment generating $\vec{Z}(\vec{\chi};t)$ obtained by taking the Fourier transform of each of the probabilities $P_A(n_h,n_c;t)$ and $P_B(n_h,n_c;t)$ with respect to the step totals $n_h$ and $n_c$. Following an approach analogous to that of Refs. \cite{dviraQAR} and \cite{modular}, we introduce the counting field $\vec{\chi}=(\chi_h,\chi_c)^T$ whose two components track the horizontal and cyclic steps, respectively. The elements of $\vec{Z}(\vec{\chi};t)$ are given by, for $\nu=A,B$,
\begin{equation}\label{eq:MGF_def}
    Z_\nu(\vec{\chi};t) = \sum_{n_h,n_c=-\infty}^\infty P_\nu(n_h,n_c;t)e^{in_h\chi_h}e^{in_c\chi_c}.
\end{equation}
Combining Eqs.~(\ref{eq:ME}) and (\ref{eq:MGF_def}), we obtain a set of differential equations for the moment generating function, $\vec{Z}$. Expressed as a matrix-vector product,
\begin{equation}\label{eq:tilted}
    \dot{\vec{Z}}(\vec{\chi};t) = W(\vec{\chi})\vec{Z}(\vec{\chi};t),
\end{equation}
with $W(\vec{\chi})$ given by,
\begin{widetext}
\begin{equation}\label{eq:tilted_matrix}
    \left(\begin{array}{cc}
        -(r_A^++r_A^-+k_B)+r_A^+e^{i(\chi_h+\chi_c)}+r_A^-e^{-i(\chi_h+\chi_c)} & k_A \\
        k_B & -(r_B^++r_B^-+k_A)+r_B^+e^{i(\chi_h-\chi_c)}+r_B^-e^{-i(\chi_h-\chi_c)}
    \end{array}\right).
\end{equation}
\end{widetext}
Eq.~(\ref{eq:tilted}) may be used to model the time evolution of $\vec{Z}$, or it may be used to directly obtain the scaled cumulant generating function (CGF) in the long time limit. This is the function whose derivatives with respect to the counting fields at $\vec{\chi}=0$ give the scaled cumulants of the associated current.
The CGF is defined as
\begin{equation}
    G(\vec{\chi}) = \lim_{t\rightarrow\infty}\frac{1}{t}\ln\sum_{\nu=A,B}Z_\nu(\vec{\chi};t),
\end{equation}
and equivalently given by the dominant eigenvalue of $W(\vec{\chi})$: that whose real part goes to zero in the limit that $\vec{\chi}\rightarrow0$ \cite{schaller}. This function is particularly useful for studying our model in the long time limit, as, due to the infinite extent of the chains, the state populations themselves do not converge to steady-state values, but the scaled cumulants (the time derivatives of the cumulants of the distributions over $n_h$ and $n_c$) do. We get,
\begin{widetext}
\begin{align}\label{eq:cgf}
    G(\vec{\chi}) = \frac{1}{2}\bigg\{
    -k_A &- k_B - r_A^+(1-e^{i(\chi_h+\chi_c)}) - r_A^-(1-e^{-i(\chi_h+\chi_c)})
    - r_B^+(1-e^{i(\chi_h-\chi_c)})-r_B^-(1-e^{-i(\chi_h-\chi_c)})
    \nonumber\\
    +\bigg(& \left[k_A + k_B + r_A^+(1-e^{i(\chi_h+\chi_c)}) + r_A^-(1-e^{-i(\chi_h+\chi_c)})+r_B^+(1-e^{i(\chi_h-\chi_c)})-r_B^-(1-e^{-i(\chi_h-\chi_c)})\right]^2
    \nonumber\\
    &-4\left[ 
    r_A^+(1-e^{i(\chi_h+\chi_c)}) + r_A^-(1-e^{-i(\chi_h+\chi_c)})][r_B^+(1-e^{i(\chi_h-\chi_c)})+r_B^-(1-e^{-i(\chi_h-\chi_c)})\right]
    \nonumber\\
    & +4k_B\left[(r_B^+ + r_B^-)(\cos(\chi_h-\chi_c)-1) + i(r_B^+-r_B^-)\sin(\chi_h-\chi_c)\right]
    \nonumber\\
    &+4k_A\left[(r_A^+ + r_A^-)(\cos(\chi_h+\chi_c)-1) + i(r_A^+-r_A^-)\sin(\chi_h+\chi_c)\right]
    \bigg)^{1/2} \bigg\}.
\end{align}
\end{widetext}
The scaled cumulants of each current can be calculated from the CGF as
\begin{equation}\label{eq:cumulants_def}
    \mathcal{C}^\mu_k=\la\la J_\mu^k\ra\ra\equiv\lim_{t\rightarrow\infty}\frac{\la\la n_\mu^k(t)\ra\ra}{t}=\frac{\partial^k}{\partial (i\chi_\mu)^k}G(\vec{\chi})\bigg|_{\vec{\chi}=0},
\end{equation}
where $\mu=h,c$.
In the next section, we study cumulants of both horizontal (symmetric) and cyclic (asymmetric) currents and analyze the information they provide on the double-chain model. 

\section{Structure and statistics}\label{sec:cumulants}
Based on Eqs.~(\ref{eq:cgf}) and (\ref{eq:cumulants_def}), it is possible to derive fully general expressions for the scaled cumulants of horizontal and cyclic currents in our model. These, in turn, depend on the parameters of the model (i.e., transition rate constants), and thus, encode information about the underlying structure of the random walk.  Given any system with a finite set of parameters and an infinite number of cumulants, one may expect to recover all of the system parameters unless either symmetry or accidental cancellations render the cumulants dependent on only a subset of parameters. This can happen if the number of cumulants that are functionally independent of each other is smaller than the number of parameters needed to describe the system. Below, we show an example of such a procedure, and how the number of cumulants needed to be computed can be higher than the number of parameters due to interdependence of some of the cumulants. 

The two lowest-order scaled cumulants of the horizontal current, representing the mean flux and scaled variance,  are given by
\begin{widetext}
\begin{align}\label{eq:horizontal_c1c2}
    \mathcal{C}_1^h &= \frac{1}{k_A + k_B}(k_A(r_A^+ - r_A^-) + k_B(r_B^+ - r_B^-)),\nonumber\\
    \mathcal{C}_2^h &= \frac{1}{k_A + k_B}(k_A(r_A^+ + r_A^-) + k_B(r_B^+ + r_B^-)) + 2\frac{k_Ak_B}{(k_A+k_B)^3}((r_A^+ - r_A^-) - (r_B^+ - r_B^-))^2.
\end{align}
Those of the cyclic current are given by
\begin{align}
    \mathcal{C}_1^c &= \frac{1}{k_A + k_B}(k_A(r_A^+ - r_A^-) - k_B(r_B^+ - r_B^-)),\nonumber\\
    \mathcal{C}_2^c &= \frac{1}{k_A + k_B}(k_A(r_A^+ + r_A^-) + k_B(r_B^+ + r_B^-)) + 2\frac{k_Ak_B}{(k_A+k_B)^3}((r_A^+ - r_A^-) + (r_B^+ - r_B^-))^2.
\end{align}
\end{widetext}
Given the expressions for one of the two currents, those for the other can be obtained by substituting $r_B^+$ for $r_B^-$ each time it appears, and vice-versa, reflecting the opposing effect that steps along chain $B$ have on each integrated current. This pattern extends to higher-order cumulants as well. 


\subsection{Example: Extracting model parameters from the cumulants}

To further illustrate the encoding of underlying structure in the cumulants, we consider a special case of a pair of chains at equilibrium ($r_A^+=r_A^-\equiv r_A$ and $r_B^+=r_B^-\equiv r_B$) and with equal rates to switch between them in either direction ($k_A=k_B=k$). This model is characterized by three parameters, so we seek out three independent expressions for the cumulants to express all of the information about the underlying chains. 
The cumulant generating function, Eq.~(\ref{eq:cgf}), simplifies in this case to a form that is symmetric under swaps of the counting fields $\chi_h$ and $\chi_c$. As a result, the cumulants of the horizontal and cyclic currents are the same. Also, due to the equilibrium condition, all odd-ordered cumulants vanish. The second cumulant simplifies to $\mathcal{C}_2^{\mu,eq} = r_A + r_B$, for $\mu=h,c$. Higher order cumulants can be obtained from Eq.~(\ref{eq:cgf}):
\begin{align}
    \mathcal{C}_4^{\mu,eq} &= 3\frac{(r_A-r_B)^2}{k} + r_A + r_B\nonumber\\
    \mathcal{C}_8^{\mu,eq} &= 63\frac{(r_A - r_B)^2}{k} - 315\frac{(r_A - r_B)^4}{k^3} + r_A + r_B.
\label{eq:C48}
\end{align}
We skip over $\mathcal{C}_6^{\mu,eq}$ above as,  surprisingly, we found that \old it is dependent on the lower order cumulants: $\mathcal{C}_6^{\mu,eq}=5\mathcal{C}_4^{\mu,eq} - 4\mathcal{C}_2^{\mu,eq}$, and thus does not contain any additional information \old. As such, one must go as high as the 8$^{\textrm{th}}$ scaled cumulant to fully characterize this system (up to the sign of $r_A-r_B$, for which an arbitrary choice can be made, due to a symmetry of the model). The three parameters can, in turn, be expressed as functions of $\mathcal{C}_2^{\mu,eq}$, $\mathcal{C}_4^{\mu,eq}$, and $\mathcal{C}_8^{\mu,eq}$, amounting to a means by which experimental data can be used to characterize the underlying features of a physical system represented by this model.  Manipulating Eq. (\ref{eq:C48}), along with the expression for the second cumulant, we get the nonintuitive results (assuming $r_A>r_B$),
\bea
k&=&\frac{35(\mathcal {C}_2^{\mu,eq}-\mathcal {C}_4^{\mu,eq})^2}{-20\mathcal {C}_2^{\mu,eq} + 21 \mathcal {C}_4^{\mu,eq} -\mathcal {C}_8^{\mu,eq}}
\nonumber\\
r_{A,B}&=&\frac{1}{2}\mathcal {C}_2^{\mu,eq} \pm \frac{1}{2}
\sqrt \frac{35(\mathcal {C}_2^{\mu,eq}-\mathcal {C}_4^{\mu,eq})^3}{-60\mathcal {C}_2^{\mu,eq}+63\mathcal {C}_4^{\mu,eq} -2\mathcal {C}_8^{\mu,eq}}.
\nonumber\\
\eea
%
These expressions are to be contrasted, e.g., by the process of diffusion on a single 
one-dimensional chain. In this case, 
 at equilibrium, the only parameter is $r_A$, and it is revealed simply through the relation $\mathcal{}{C}_2=2r_A$. Even in the presence of a bias, \old
$\mathcal{C}_1=r_A^+-r_A^-$ and $\mathcal{C}_2=r_A^++r_A^-$, so
knowledge of the first two cumulants suffices to extract the forward and reversed rates.

We stress here that while the two currents in the system have the exact same statistics at equilibrium, their cumulants still reflect the fact that there are three distinct parameters characterizing the system, and therefore behave differently from those of a simple, one-dimensional random walk.

For instance, without carrying out the analysis of higher-order cumulants, the absence of mean fluxes means it may be difficult to distinguish this model from a simple random walk on a single chain at equilibrium. Our model is equivalent to a one-dimensional random walk only when setting $r_A=r_B$, leading to a situation in which there is no meaningful distinction between chains $A$ and $B$, all even cumulants become equal to $\mathcal{C}_2^{h,eq}$, and analysis of the horizontal current cannot, in turn, reveal any information about the inter-chain switching rate, $k$. 
Outside of this equal-rate case, however, it is only when looking to the higher-order cumulants that one can determine if there are, indeed, two distinct chains facilitating transport in the horizontal direction (e.g., one characterized by faster transitions and one characterized by slower transitions). The overall probability distribution for this model differs from that of a simple, one-dimensional random walk even at equilibrium, but this deviation is revealed only if one looks beyond the level of the variance.

The arguments presented here in reference to the equilibrium case may extend to greater levels of generality captured by the cumulant generating function, Eq.~(\ref{eq:cgf}). More general cases correspond to models described by a larger number of parameters. As a result, one has to look to more cumulants of the currents to have enough independent equations to fully characterize the model.

\subsection{Illustrative limiting behaviors}
In addition to the cases of equilibrium, 
one can consider different interesting limits characterizing particular behaviors of the model to further illustrate the relationship the underlying structure we consider here has with the statistics of currents.

In one such limit, one of the chains, say chain $B$, permits no horizontal transitions: that is, $r_B^+=r_B^-=0$. In this case the model may be considered a random walk on a one-dimensional chain (chain $A$), but with a one-site-long ``side chain'' at each site. It can be shown that the cumulants of the
symmetric ($h$) and asymmetric ($c$) currents are equal, since the only contributions to either are horizontal steps along chain $A$. 
The first two cumulants become, for $\mu=h,c$,
\begin{align}\label{eq:rb_zero}
    \mathcal{C}_1^\mu\big|_{r_B^+=r_B^-=0} =& \frac{k_A}{k_A+k_B}(r_A^+-r_A^-),\nonumber\\
    \mathcal{C}_2^\mu\big|_{r_B^+=r_B^-=0} =& \frac{k_A}{k_A+k_B}(r_A^++r_A^-)\nonumber\\
    &+ 2\frac{k_Ak_B}{(k_A+k_B)^3}(r_A^+ - r_A^-)^2.
\end{align}
The first cumulant very closely mimics that of a random walk on a single chain, but with the addition of a prefactor $k_A/(k_A+k_B)$, which quantifies the proportion of the total time that the walker spends on the $A$ chain, where it has horizontal mobility. In this regard, time spent in the $B$ sites can be understood as occasional ``pauses'' taking place during a 1-d random walk. This ability to pause has further effects on the second cumulant, whose first term above mimics a scaled-down equivalent to a 1-d random walk, but whose second term reflects the fact that the additional structure provided by the $B$ sites can lead to an enhancement of the variance. This is the case only when $r_A^+\neq r_A^-$, i.e., transport along the $A$ chain is {\it biased}, as the pausing effect can be understood as increasing the likelihood of the walker lagging behind when a bias would otherwise have the entire population distribution shift in one direction as time passes. This limit could be a starting point for examining random walks with larger or topologically nontrivial side chains.

Another limit of interest is that of high bias, i.e., the unidirectional case, in which $r_A^-=r_B^-=0$. A random walk on a single chain in this case is characterized by Poisson statistics, wherein all scaled cumulants take on equal values. The possibility of hops between chains $A$ and $B$, however, lead to deviations from this. For instance, there is, once again, enhancement of the second cumulant as compared to the single-chain case. Take, as an example, the second cumulant of the horizontal current; this quantity becomes
\begin{align}\label{eq:unidirectional}
    \mathcal{C}_2^h\big|_{r_A^-=r_B^-=0} = \frac{k_Ar_A^+ + k_Br_B^+}{k_A + k_B}
    +2\frac{k_Ak_B}{(k_A+k_B)^3}(r_A^+-r_B^+)^2.
\end{align}
The first term is precisely equal to $\mathcal{C}_1^h$ in this limit. However, the second term, which depends on the inter-chain transition rate constants even in the case that $k_A=k_B=k$, reflects the additional uncertainty in the position of the walker associated with the occasional switches between chains. 
In this case, as in the aforementioned case where horizontal transitions are only permitted on one chain, measurements of the second cumulant can be used to probe the underlying network structure, distinguishing the behavior of this system from that of a simple one-dimensional random walk in an analogous limit.

Finally, we consider the above limits in the case that both $k_A$ and $k_B$ are very large relative to all other rates. In this case, the second term in the second cumulant vanishes for both Eqs.~(\ref{eq:rb_zero}) and (\ref{eq:unidirectional}). Very fast transitions between the two chains mean that the horizontal current can be treated accurately as that of a one-dimensional random walk with properties simply averaged between the two chains. In other words, the horizontal transitions are still described accurately as a Markov process even after coarse-graining by grouping together the pairs of $A$ and $B$ sites at each position.

\section{Zero-current noise and deviation from equilibrium}\label{sec:zc_and_entropy}

The complex dependence of the cumulants of horizontal and cyclic currents on the parameters of the underlying model may be useful in assessing the degree to which the system deviates from equilibrium.

This deviation is quantified by the entropy production rate for the system. To calculate this quantity, we assume that individual jumps along the $A$ and $B$ chain are associated with energy changes that are uniform for all pairs of adjacent sites along each chain, but which may vary between the two chains. These are denoted $\Delta_A$ and $\Delta_B$, respectively. Correspondingly, a jump downwards in energy along one chain leads to the dissipation of energy into a bath maintained at inverse temperature $\beta$. Considering the case where $k_A=k_B=k$, the inter-chain transitions can be interpreted as being coupled to an infinite-temperature bath, or some non-thermal switching process, and are, in either case, associated with no contributions to the entropy production.

Given the energy splittings $\Delta_\nu$ ($\nu=A,B$), it is possible to relate forward and reverse transition rates along each chain via detailed balance relations,
\begin{equation}\label{eq:db}
    \frac{r_\nu^-}{r_\nu^+} = e^{-\beta\Delta_\nu},
\end{equation}
where we have identified $\beta\Delta_\nu$ as the quantity of entropy produced in a rightward 
jump along the $\nu$ chain. The case of zero horizontal current demands that $\Delta_A$ and $\Delta_B$ have opposing signs.

In the steady state, the overall mean entropy production rate is given by the sum of thermodynamic force-flux products,
\begin{equation}\label{eq:epr_def}
    \la\sigma\ra = \sum_{\nu=A,B}\beta\Delta_\nu\mathcal{C}_1^\nu,
\end{equation}
where we have invoked the first scaled cumulant for the current along each individual chain. $\mathcal{C}_1^A$ and $\mathcal{C}_1^B$ can be computed from Eq.~(\ref{eq:cgf}) with the variable substitution $(\chi_h,\chi_c)\rightarrow(\chi_A,\chi_B)$, defined as,
\begin{equation}
    \chi_A = \frac{\chi_h + \chi_c}{2};\:\:\:\:\:\:\chi_B = \frac{\chi_h - \chi_c}{2}.
\end{equation}
The resulting mean fluxes are
\begin{align}
    \mathcal{C}_1^A =& \frac{1}{2}\left(\mathcal{C}_1^h + \mathcal{C}_1^c\right) = \frac{k_A}{k_A + k_B}\left(r_A^+ - r_A^-\right)\nonumber\\
    &\:\xrightarrow{k_A=k_B}\frac{1}{2}r_A^+\left(1-e^{-\beta\Delta_A}\right)\nonumber\\
    \mathcal{C}_1^B =& \frac{1}{2}\left(\mathcal{C}_1^h - \mathcal{C}_1^c\right) = \frac{k_B}{k_A + k_B}\left(r_B^+ - r_B^-\right)\nonumber\\
    &\:\xrightarrow{k_A=k_B}\frac{1}{2}r_B^+\left(1-e^{-\beta\Delta_B}\right).
\end{align}
This yields, from Eq.~(\ref{eq:epr_def}), 
\be\label{eq:epr}
    \la\sigma\ra = \frac{\beta\Delta_A}{2}r_A^+\left(1-e^{-\beta\Delta_A}\right) + \frac{\beta\Delta_B}{2}r_B^+\left(1-e^{-\beta\Delta_B}\right).
\ee
The entropy production rate is (trivially) zero when the system is unbiased, $\Delta_A=\Delta_B=0$. 

\subsection{Zero-current noise}

One particular situation of interest is the ``zero-current" case, in which the first cumulant of the horizontal current vanishes, but the system is not necessarily at equilibrium and $\mathcal{C}_1^c$ may take on a finite value. Considering the case that $k_A=k_B=k$, we return to Eq.~(\ref{eq:horizontal_c1c2}) and identify the condition for $\mathcal{C}_1^h$ to vanish as
\be\label{eq:zero_current1}
    r_A^+ - r_A^- = r_B^- - r_B^+.
\ee
This condition, re-expressed in terms of the energy changes, is
\be
r_A^+(1-e^{-\beta\Delta_A}) = r_B^-(1-e^{\beta\Delta_B}).
\label{eq:zerocurrent}
\ee
If an observer has access to information only about this horizontal current, then the vanishing mean alone would not provide any evidence that the system is away from equilibrium, even though, as is clear from Eq.~(\ref{eq:epr}), the entropy production rate may very well be nonzero. The analogous zero-cyclic-current case can also be considered.

Of course, as is clear from Eq.~(\ref{eq:epr_def}), if one is calculating the entropy production rate exactly, they must have knowledge of both independent mean currents $\mathcal{C}_1^A$ and $\mathcal{C}_1^B$, which implies they have access to information about the cyclic and horizontal currents. If the cyclic currents cannot be measured, one cannot rely on an exact calculation of $\la\sigma\ra$, and instead needs to turn to the results of Sec. \ref{sec:cumulants} to determine through the values of higher-order cumulants of the horizontal current whether or not the system is at equilibrium.

The second cumulant (i.e. noise) can indicate that the system is out-of-equilibrium, despite zero current observed. The simplest case of this is when $\Delta_A = -\Delta_B$ and $r_A^+ = r_B^-$. 
The horizontal noise in Eq. (\ref{eq:horizontal_c1c2}) then simplifies to
\begin{align}
    \mathcal{C}_2^h &= 
    (r_A^++r_A^-) + \frac{1}{k}\left(r_A^+-r_A^-\right)^2
\end{align}
The two terms in $\mathcal{C}_2^h$ represent contributions from equilibrium thermal noise and out-of-equilibrium shot noise. 

\begin{figure}
    \centering
    \includegraphics[scale=0.4]{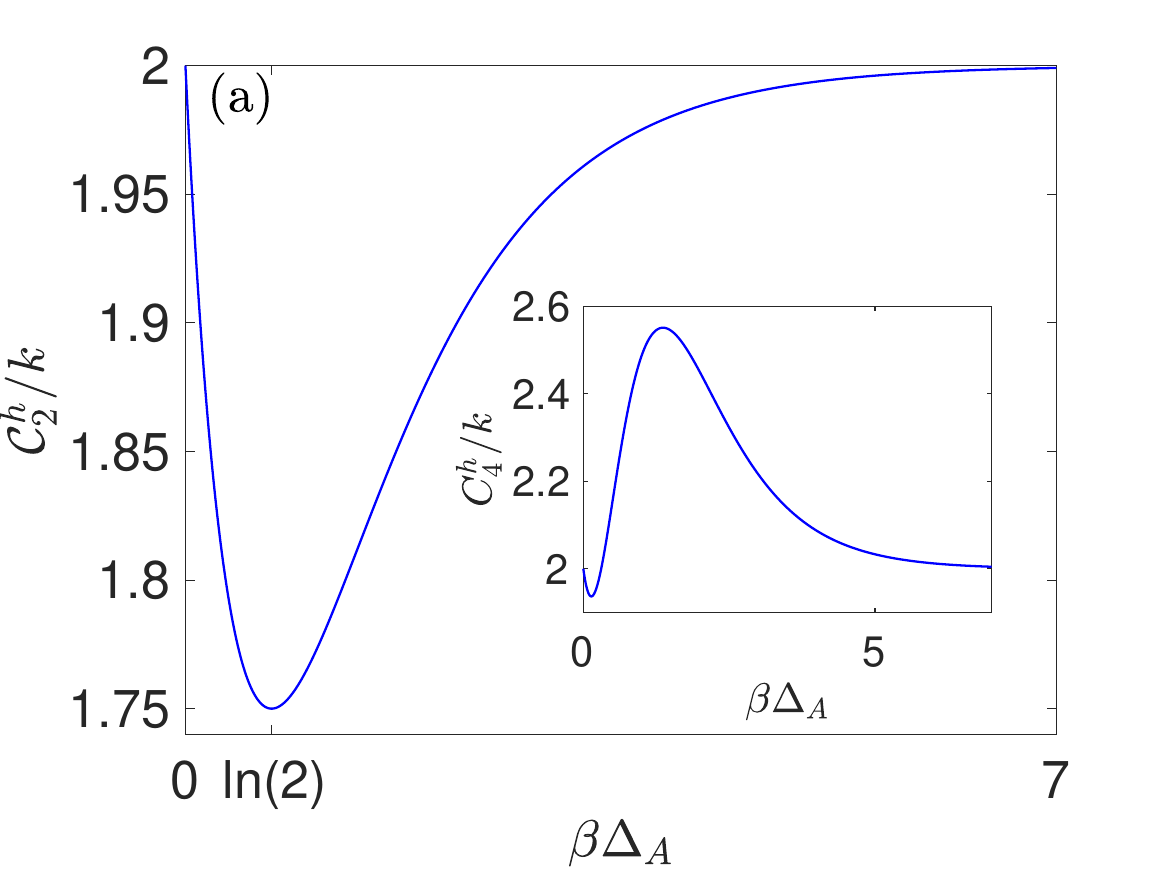} 
    \includegraphics[scale=0.4]{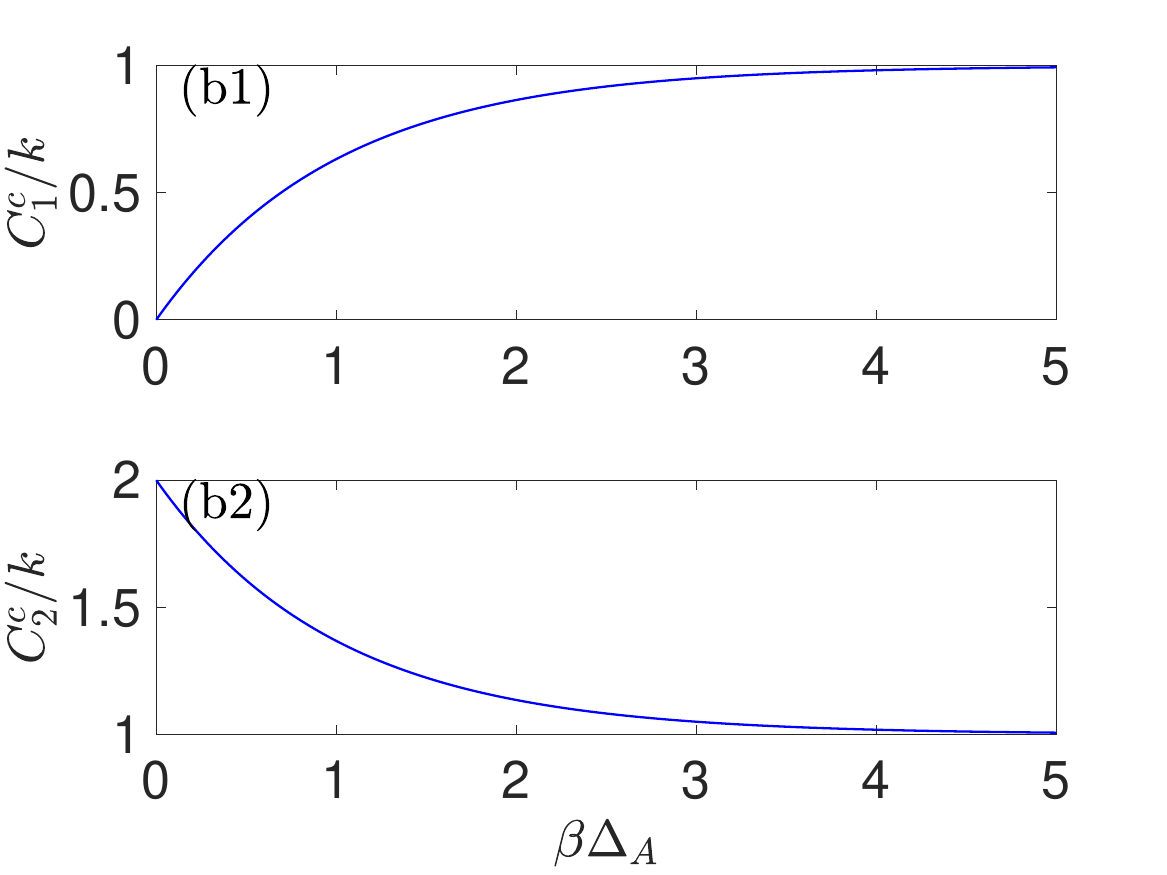}
    \caption{(a) The scaled zero-current noise $\mathcal{C}_2^h/k$ under the condition of zero {\it horizontal} current, $r_A^+ = r_B^-$, varying with $\Delta_A = -\Delta_B$, and setting $r_A^+=k$. The scaled $\mathcal{C}_4^h/k$ is also presented (inset), with the odd cumulants being zero. (b1)-(b2) The first and second scaled {\it cyclic} cumulants $\mathcal{C}_1^c/k$ and $\mathcal{C}_2^c/k$ varying with $\Delta_A = -\Delta_B$.} 
    \label{fig:noise}
\end{figure}

Fig. \ref{fig:noise} plots cumulants at varying values of $\beta \Delta_A$, where we furthermore set $r_A^+=r_B^-=k$. 
It is interesting to point out that in this case, the second cumulant $\mathcal{C}_2^h$ is the same at the extreme bias $\Delta_A\rightarrow \infty$ limit, and at zero bias. This should be understood as follows. At extreme bias, the motion along the legs of the ladder is essentially deterministic, with the only probabilistic element coming from the walker randomly hopping between the upper and lower legs of the ladder, which maps to a simple balanced random walk. On the other hand, at zero bias, the system is completely symmetric between the upper and lower sides. This case, too, is described by a balanced random walk, since the hops between the upper and lower legs do not affect the statistics of lateral motion. Since we expect the high/low bias limits to behave the same, we expect the second cumulant $\mathcal{C}_2^h$ to be extremized for some intermediate choice of  $\Delta_A$ between these limits. Indeed, we find such an extremum at $\beta \Delta_A = \ln 2$, where $\mathcal{C}_2^h = \frac{7}{4}k$. Curiously, the second cumulant is actually minimal, rather than maximal at this point.
 Rather, it is maximized at the unbiased limit and at extremely large bias, where it approaches $2k$. Similarly, for $\mathcal{C}_4^h$, the equilibrium and far-from-equilibrium values are identical ($2k$), and a peak is observed at the extremum. 

To discern the limits of high and low bias, we consider the first and second {\it cyclic} cumulants.  Indeed, the cyclic cumulants reveal further information, not given by the statistics of overall lateral motion, as discussed above. 
Under the same conditions, the first and second cyclic cumulants are
\begin{align}
    \mathcal{C}_1^c &= k \left( 1-e^{-\beta \Delta_A} \right),  \nonumber\\
    \mathcal{C}_2^c &= k \left( 1+e^{-\beta \Delta_A} \right).
    \label{eq:cyc-cur}
\end{align}
As plotted in Fig. \ref{fig:noise}, these cumulants reveal that the system is out-of-equilibrium. At large $\beta \Delta_A$, their values tend to $k$.
%

\begin{figure}[htbp]
    \includegraphics[scale=0.4]{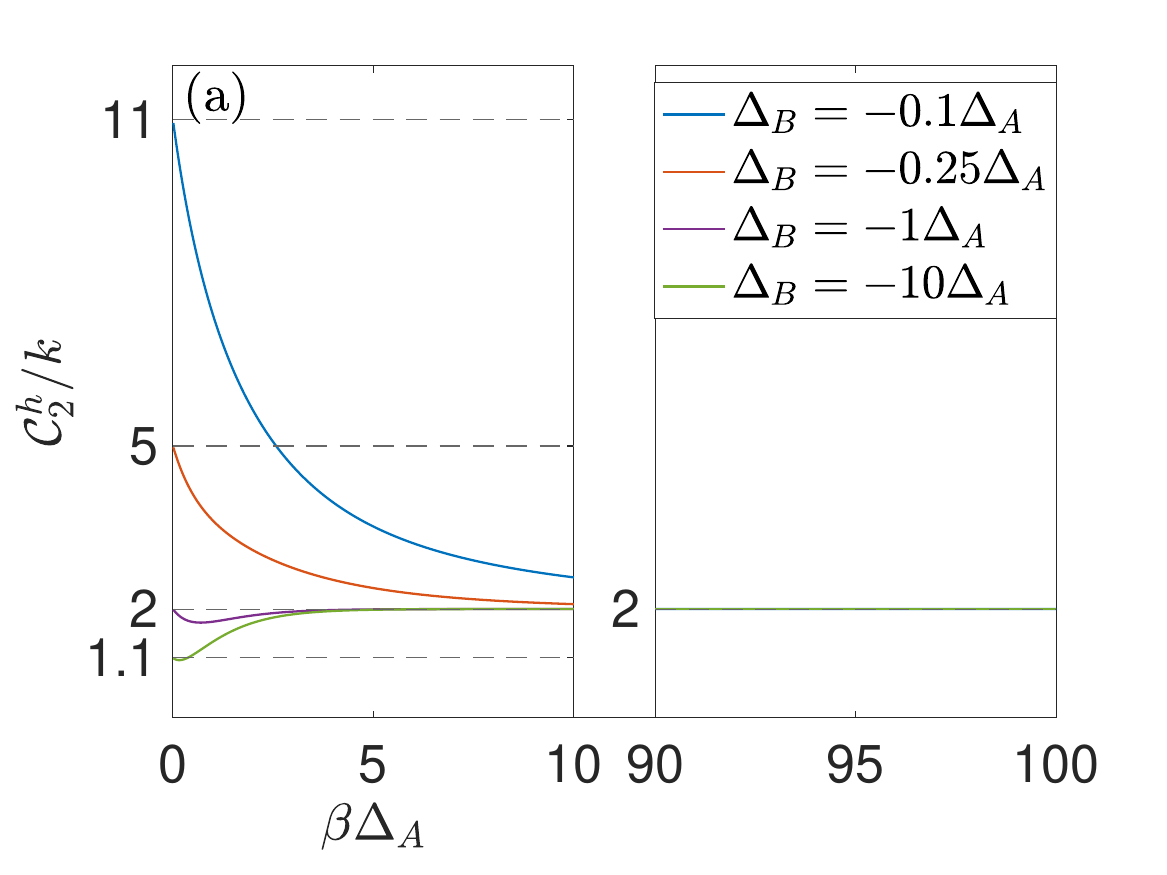}
    \includegraphics[scale=0.4]{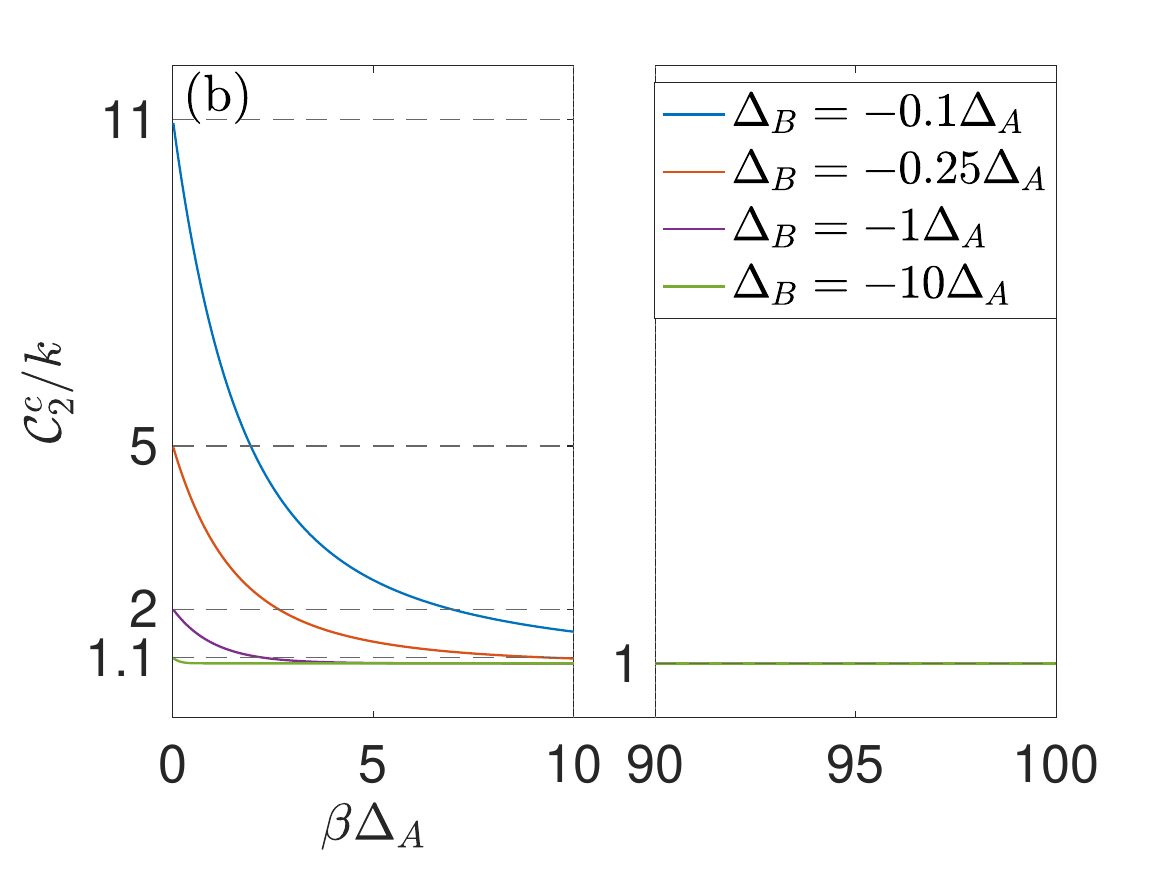}
    \caption{The scaled zero-current noise under the general condition $\Delta_A \neq -\Delta_B$. We present (a) 
         $\mathcal{C}_2^h/k$ and (b) $\mathcal{C}_2^c/k$ setting for simplicity $r_A^+ = k$ . Note that the purple lines are identical to Fig. \ref{fig:noise}. Dashed lines indicate asymptotic values as $\beta \Delta_A \rightarrow 0$. At large $\beta \Delta_A$, $\mathcal{C}_2^h \rightarrow 2k$ and $\mathcal{C}_2^c \rightarrow k$. The horizontal axes are discontinuous, such that each figure probes both the low- and high-bias regimes.} 
    \label{fig:noise-general}
\end{figure}

In the most general zero-current case, $\Delta_A \neq -\Delta_B$. Keeping for simplicity $r_A^+ = k$, the horizontal second cumulant is now
\begin{align}
    \mathcal{C}_2^h &= k \left[ \frac{e^{-\beta \Delta_B}-e^{-\beta \Delta_A}}{e^{-\beta \Delta_B} - 1} + (1-e^{-\beta \Delta_A})^2 \right].
\end{align}
Varying $\beta \Delta_B$ as multiples of $\beta \Delta_A$, we obtain Fig. \ref{fig:noise-general}. Expanding the expression for $\mathcal{C}_2^h$ at small $\beta \Delta_\nu$ to first order, the scaled horizontal cumulant approaches $\mathcal{C}_2^h/k = 1-\Delta_A/\Delta_B$, distinct from the equilibrium value of 2. Thus, at low biases, fluctuations of the horizontal current indicate that the system is out-of-equilibrium despite the zero-current condition. At large biases, the horizontal second cumulant approaches $2k$. For the cyclic cumulants, $\mathcal{C}_1^c$ is identical to Eq.~(\ref{eq:cyc-cur}). The cyclic second cumulant is simply the first term of the horizontal second cumulant: 
\begin{align}
    \mathcal{C}_2^c &= k \left( \frac{e^{-\beta \Delta_B}-e^{-\beta \Delta_A}}{e^{-\beta \Delta_B} - 1} \right).
\end{align}
Similar to the scaled horizontal noise, this expression approaches $\mathcal{C}_2^c/k = 1-\Delta_A/\Delta_B$ at low biases. However, at large bias it reaches $k$, contrasting the equilibrium-like behavior of $\mathcal{C}_2^h$ at large bias.
Fig. \ref{fig:noise-general} presents the second cumulant of both horizontal and cyclic currents, demonstrating the above limits. 

In this section, we focused on the case where the horizontal current is zero, but the cyclic current reveals information about the underlying non-equilibrium nature of the model. The analogous case where the cyclic current is zero (and the horizontal current is nonzero) is complimentary, and one can derive the same results under the exchange of superscripts $h$ with $c$ and $\Delta_B$ with $-\Delta_B$. 

\subsection{Lower boundary on the entropy production rate} 
Information about the cumulants of a current may, in some cases, also be used to lower bound the entropy production rate via the thermodynamic uncertainty relation (TUR). This relation bounds the entropy production with respect to the precision of any steady-state current, taking the form \cite{udoTUR, ging16, ging17, li2019, horowitz2020, bijayTUR_hierarchy, kamijima2023},
\begin{equation}\label{eq:tur}
    \frac{\la \sigma\ra}{2}\geq\frac{(\mathcal{C}_1^\mu)^2}{\mathcal{C}_2^\mu}.
\end{equation}
For the model in question, we may set $\mu=h,c$. One can alternatively write down TUR bounds based on the currents along individual chains, $A$ or $B$, though we note that there are only two independent bounds for our model due to the relationships between the currents.

The TUR offers a mechanism to estimate the entropy production rate that requires only the two lowest-order cumulants of a given current, while an exact calculation requires knowledge of four independent parameters, as per Eq.~(\ref{eq:epr}). Again, this may be useful when information about one current is hidden, though we note that if the mean value of the observable current is zero, the TUR bound is satisfied trivially. Other approaches towards estimating the entropy production rate when some information is hidden include optimizing over underlying structures consistent with observed statistics, and considering wait time distributions \cite{martinez2019, skinner1, skinner2, nitzan2023}.

\begin{figure}[h]
    \centering
    \includegraphics[width=0.9\columnwidth]{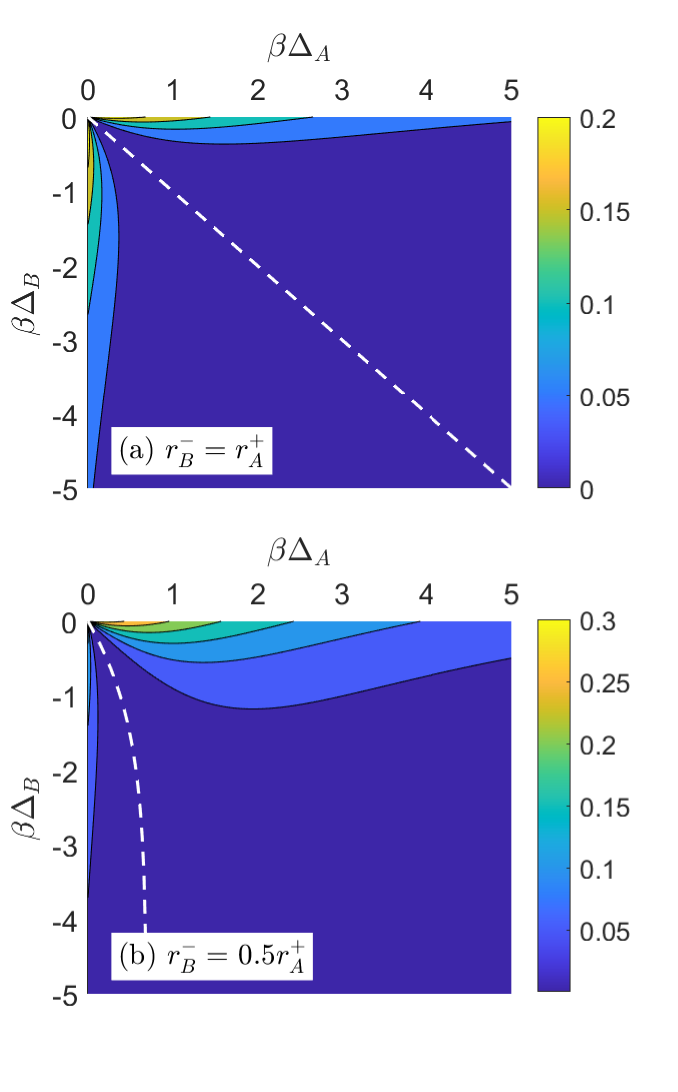}
    \caption{Analysis of the tightness of the TUR bound.
    The quantity $(1/\la\sigma\ra)(\mathcal{C}_1^h)^2/\mathcal{C}_2^h$ is plotted, while varying $\Delta_A$ and $\Delta_B$. Note that $r_B^-$ is fixed at (a) $r_A^+$ and (b) $0.5r_A^+$. $r_B^+$ then varies with $\Delta_B$ via Eq~(\ref{eq:db}). In most of the parameter regime considered, this quantity does not come near its maximum value of 0.5, particularly near the zero-current condition (indicated by the dashed white line), where the TUR bound is trivially satisfied. $r_A^+ = k = 1$.}
    \label{fig:turh_contour}
\end{figure}

\begin{figure}[h]
    \centering
    \includegraphics[width=0.9\columnwidth]{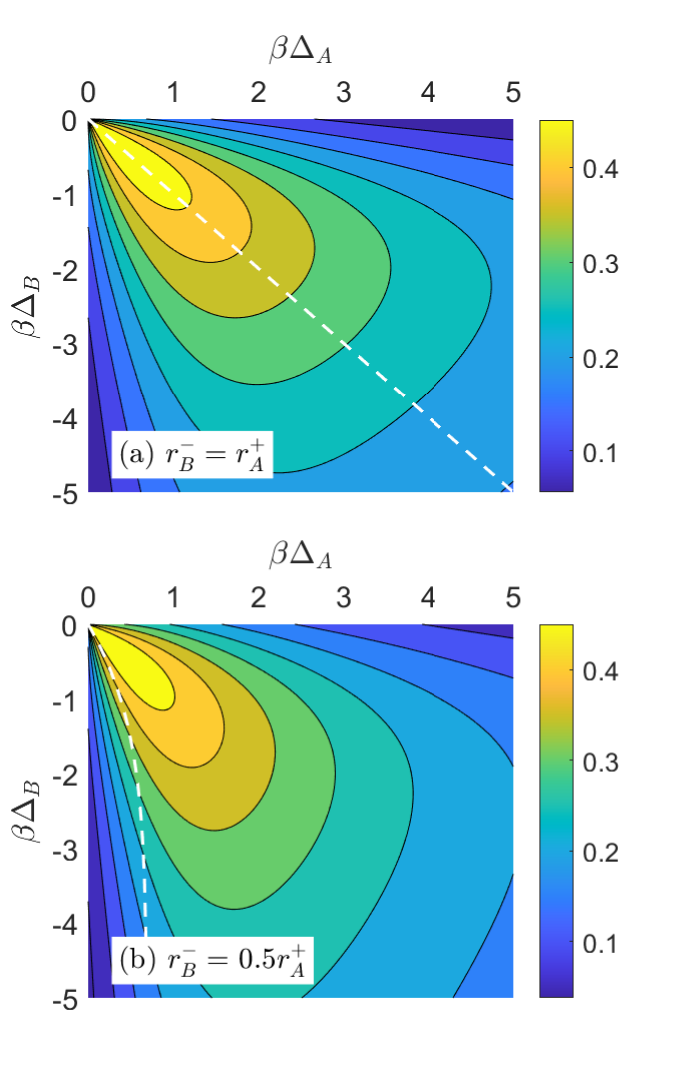}
    \caption{
    Analysis of the tightness of the TUR bound.
    The quantity $(1/\la\sigma\ra)(\mathcal{C}_1^c)^2/\mathcal{C}_2^c$, varying with $\Delta_A$ and $\Delta_B$, at (a) $r_B^-=r_A^+$ and (b) $r_B^-=0.5r_A^+$. The condition of zero \textit{horizontal} current is indicated by the dashed white line. In most of the parameter regime studied, the cyclic current takes on higher mean values than its horizontal counterpart, enabling much nearer saturation of the TUR, particularly close to equilibrium. $r_A^+ = k = 1$.}
    \label{fig:turc_contour}
\end{figure}

The tightness of the TUR is examined in Figs. \ref{fig:turh_contour} and \ref{fig:turc_contour}, where the quantity $(1/\la\sigma\ra)(\mathcal{C}_1^\mu)^2/\mathcal{C}_2^\mu$ is plotted while varying both $\Delta_A$ and $\Delta_B$. In accordance with Eq.~(\ref{eq:tur}), this quantity must be between 0 and 0.5. Negative values of $\Delta_B$ are considered, as the opposing signs of $\Delta_A$ and $\Delta_B$ enable the condition of zero horizontal current, which is depicted on each plot with a dashed white line. Note that for each set of axes we fix the value of $r_B^-$, and allow $r_B^+$ to vary with the bias according to detailed balance.

The plots show that near the condition of zero horizontal current, the TUR bound associated with the horizontal current is indeed not particularly tight, and thus does not provide a great estimate of the entropy production rate. The bound associated with the cyclic current is much tighter, nearly saturating the inequality in the near-equilibrium regime, suggesting that knowledge of the two lowest cumulants of this current can provide a better estimate of the entropy production rate.

\begin{figure*}[htpb]
    \centering
    \includegraphics[width=1\textwidth]{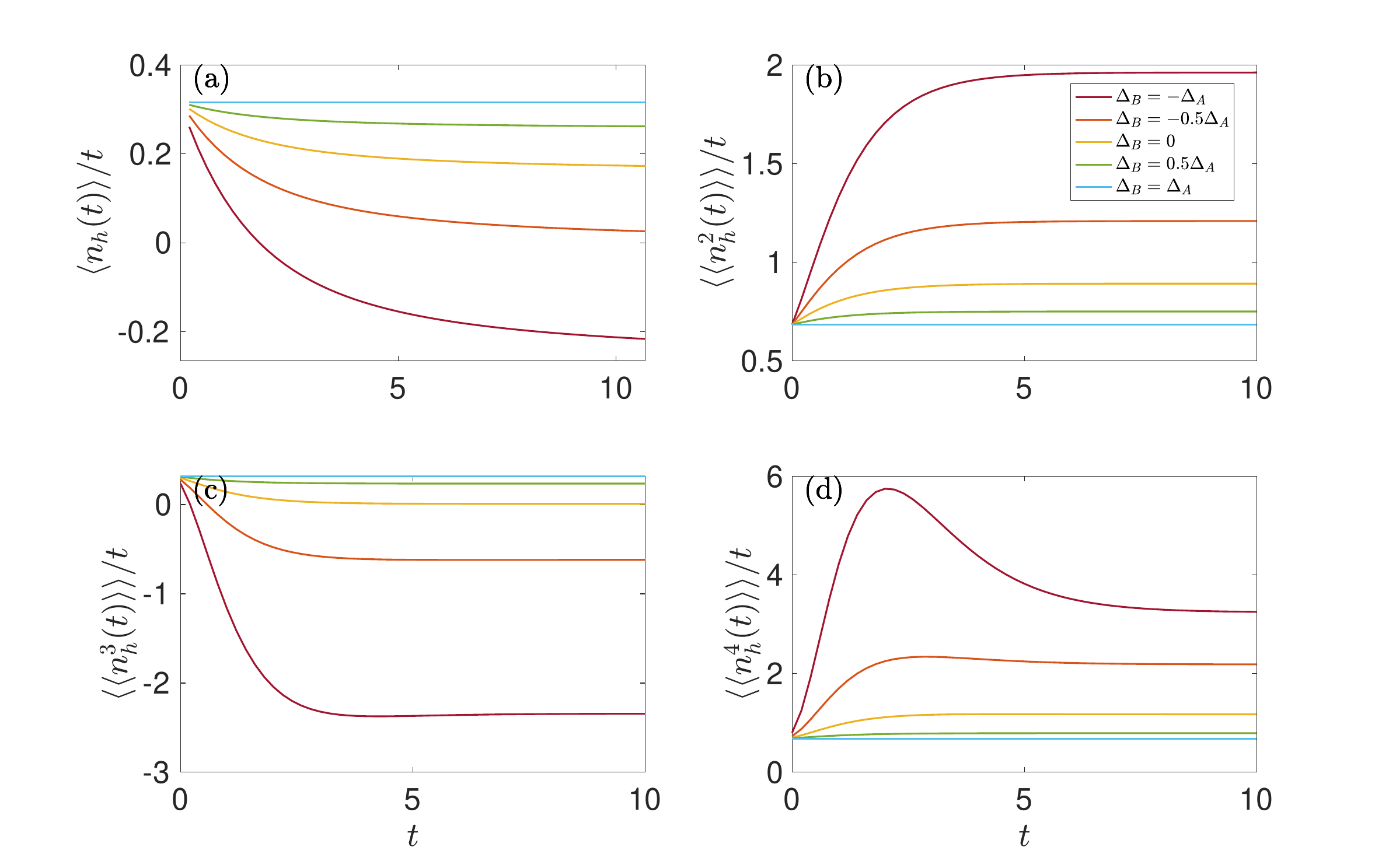}
    \caption{(a)-(d) Numerically calculated dynamics of the first four scaled cumulants 
    for different ratios of $\Delta_B/\Delta_A$. For simplicity, $r_A^+ = r_B^+ = k = 0.5$ and $\beta \Delta_A = 1$. Note that the condition that $\Delta_B = -\Delta_A$ does not guarantee zero current, since $r_A^+ \neq r_B^-$.} 
    \label{fig:dyn}
\end{figure*}
\section{Dynamics}\label{sec:dynamics}

While the steady-state scaled cumulants $\mathcal{C}_k^\mu$ are useful for studying structure, they are obtained at the long time limit. If the initial state of the system is known, certain information about its structure may also be inferred by studying the dynamics of the system as it approaches steady state.
In this section, we point out that in our model, the decay of initial conditions is governed by a single rate constant, which controls the inter-chain transition process.

To understand the dynamics of our model, we must solve the equations of motion, Eq. (\ref{eq:ME}). We will only analyze dynamics associated with the horizontal current and leave aside the cyclic current, assuming it is experimentally inaccessible. Setting $k_A = k_B = k$ and focusing on the horizontal current, the equation of motion for the probabilities to occupy chain $\mu=A,B$ after passing $n_h$ horizontal steps is given by 
\bea
\label{eq:horizontal}
    \dot{P}_A(n_h;t) 
   & = &-P_A(n_h;t)(r_A^+ + r_A^- + k) + P_B(n_h;t)k\nonumber\\
    &+& P_A(n_h-1;t)r_A^+ + P_A(n_h+1;t)r_A^- \nonumber\\
    \dot{P}_B(n_h;t)
    &= &-P_B(n_h;t)(r_B^+ + r_B^- + k)+ P_A(n_h;t)k\nonumber\\
    &+& P_B(n_h-1;t)r_B^+ + P_A(n_h+1;t)r_B^-.
    \nonumber\\
\eea
We multiply the differential equations by $n_h$ and sum both equations over all possible values $n_h \in (-\infty,\infty)$ to derive an equation of motion for the first moment:
\begin{align}\label{eq:1stmoment}
    \frac{d\la n_h(t)\ra}{dt} = &(r_A^+ - r_A^-) \sum_{n_h} P_A (n_h;t) \nonumber\\
    &+ (r_B^+ - r_B^-) \sum_{n_h} P_B (n_h;t).
\end{align}
From the equation for $\dot{P}_A(n_h;t) $ in Eq. (\ref{eq:horizontal}), summing over $n_h$, then solving, leads to
\begin{align}
    \sum_{n_h} P_A(n_h;t) = \frac{1}{2} \left(1+e^{-2kt}P_0\right),
    \label{eq:timePA}
\end{align}
where $P_0 \equiv \sum_{n_h} \left[P_A(n_h;0)-P_{B}(n_h;0)\right]$. An analogous expression can be derived for $\sum_{n_h} P_B(n_h;t)$ under the exchange of $A$ and $B$ in Eq. (\ref{eq:timePA}),
\begin{align}
    \sum_{n_h} P_B(n_h;t) = \frac{1}{2} \left(1-e^{-2kt}P_0\right).
\end{align}
Using these two sums along with Eq. (\ref{eq:1stmoment}), we solve for the scaled first moment,
\begin{align}\label{eq:C1dyn}
    \frac{\la n_h(t) \ra}{t} =& \frac{r_A^+ - r_A^- + r_B^+ - r_B^-}{2}\nonumber\\
    &+ \frac{\left( 1-e^{-2kt} \right)(r_A^+ - r_A^- - (r_B^+ - r_B^-))}{4kt}P_0.
\end{align}
%
Accordingly, the relaxation timescale depends only on $k$. As such, studying the dynamics provides a direct method for determining the rate with which the transitions between chains $A$ and $B$ occur. At large $kt$, we obtain the first cumulant as presented in Eq. (\ref{eq:horizontal_c1c2}). If we impose the zero (horizontal) current condition, we are only left with the second, time-dependent term:
\begin{align}
    \frac{\la n_h(t)\ra}{t} = \frac{\left( 1-e^{-2kt} \right)(r_A^+ - r_A^-)}{2kt}P_0.
\end{align}
The time dependence of the scaled first moment obeys a simple exponential decay. 
At large values of $k$, the system reaches steady state sooner.

However, the condition that $r_A^+ - r_A^- = r_B^+ - r_B^-$ leads to a completely time-independent expression for the scaled first moment, as reflected in Eq.~(\ref{eq:C1dyn}). 
That is, we obtain the first horizontal cumulant instantly and for all times $t$ when the cyclic current is zero. This is because the net rate at which the walker proceeds forward is the same, regardless of the chain on which it happens to dwell at any given moment. The inter-chain switching therefore has no impact on this, and there is no sense in which it takes time for the first cumulant to ``relax'' to its steady-state behavior.


We visualize the dynamics of the first four scaled cumulants 
for various ratios of $\Delta_B/\Delta_A$ in Fig. \ref{fig:dyn}. Simulations were done following the procedure used in Ref. \citenum{modular}: A finite rate matrix was constructed from Eq. (\ref{eq:ME}), with the initial condition that the walker starts on site $n=0$ on Chain A. A total of 322 sites were used, half of which belonged to each chain. The rate matrix was exponentiated to numerically solve for the probability distribution functions over time. 
The simulation time was limited to ensure the probability distributions do not meaningfully reach the boundaries.
It is interesting to note that the cumulants are not necessarily monotonic in time, as clearly observed for the kurtosis.

For $\Delta_A = \Delta_B$, steady state is immediately achieved for all cumulants, because the two chains are identical (given $r_A^+ = r_B^+$) thus resulting in the standard one-dimensional random walks. Conversely, it takes longer times to attain steady state values as $\Delta_B \rightarrow -\Delta_A$. We therefore require $\Delta_A \neq \Delta_B$ to gain information on the system's timescale and the value of $k$. Otherwise, only the steady-state value is observed.

\section{Summary}\label{sec:summary}

We studied the full counting statistics of a random walk model with two coupled chains. The model supports two independent currents, horizontal and cyclic, and it allows the examination of several basic questions concerning random walks in multicyclic networks. In the steady state regime:
(i) We showed that nontrivial combinations of cumulants allow one to identify the model parameters.
(ii) Under the condition of zero horizontal current, we showed that fluctuations of the flux, as well as the complementary cyclic current, reveal the non-equilibrium condition.
(iii) We exemplified that different fluxes have distinct capabilities for providing tighter lower bound on the entropy production rate. 
We further studied the dynamics of the system before reaching steady state, and (iv) showed that the relaxation rate to steady state only depends on the inter-chain rate constant, rather than on intrinsic parameters.

Open questions concern developing a fundamental understanding of the physical information each cumulant conveys and why certain cumulants are redundant for extracting model parameters. 
Interesting extensions of this work include investigations of fluctuations in different network topologies related to biological and chemical reaction networks and studies of related quantum network models, where coherent dynamics is impacted either by a protocol of frequent measurements \cite{Klich22,Klich23} or by incoherent effects induced by a thermal environment.  

\begin{acknowledgements}
J. Li was supported by the University of Toronto Excellence Award.
M.G. acknowledges support from the NSERC Canada Graduate Scholarship-Doctoral.  
D.S. acknowledges the NSERC discovery grant and the Canada Research Chairs Program.
\end{acknowledgements}

\end{document}